\documentclass[
    ,final            
  ]
  {aipproc}

\layoutstyle{6x9}
\usepackage{stmaryrd}
\usepackage{graphicx}
\usepackage{amssymb}
\usepackage{amsmath}
\usepackage{units}
\usepackage{url}
\usepackage{wrapfig}
\usepackage{floatrow}
\usepackage[font={bf,small},textfont={md,footnotesize},labelsep=period]{caption}
\newfloat{ffig}{thp}{log}
\floatname{ffig}{FIGURE}

\graphicspath{{Figs/}}
\DeclareGraphicsExtensions{.eps,.ps,.eps.gz,.ps.gz}
\newcommand{\pslash}{p\llap{/\kern+0.1em}}

\newcommand{\xipb}{\ensuremath{\:\textrm{pb}^{-1}}}

\newcommand{\xGeV}{\ensuremath{\:\textrm{Ge\kern -0.1em V}}}
\newcommand{\GeV}{\ensuremath{\textrm{Ge\kern -0.1em V}}}
\newcommand{\xGeVs}{\ensuremath{\:\textrm{Ge\kern -0.1em V^{2}}}}
\newcommand{\xMeV}{\ensuremath{\:\textrm{Me\kern -0.1em V}}}
\newcommand{\MeV}{\ensuremath{\textrm{Me\kern -0.1em V}}}
\newcommand{\KeV}{\ensuremath{\:\textrm{ke\kern -0.1em V}}}
\newcommand{\eV}{\ensuremath{\:\textrm{e\kern -0.1em V}}}

\newcommand{\RAPGAP}{\textsc{Rapgap}}

\begin{document}

\title{Beauty Production in Deep Inelastic Scattering at HERA using Decays into Electrons}


\author{R.~Shehzadi {\footnotesize{(for the ZEUS Collaboration)}}}{
  address={Physikalisches Institut, Universit\"at Bonn, Germany}
}

\begin{abstract}
  The results from a recent analysis on beauty production in deep
  inelastic scattering at HERA using decays into electrons from the
  ZEUS collaboration are presented. The fractions of events containing
  $b$ quarks were extracted from a likelihood fit using variables
  sensitive to electron identification as well as to semileptonic
  decays. Total and differential cross sections were measured and
  compared with next-to-leading-order QCD calculations. The beauty
  contribution to the proton structure function $F_2$ was extracted
  from the double-differential cross sections.

\end{abstract}

\maketitle


\section{Introduction}
The measurement of beauty production in $ep$ collisions at HERA
provides a powerful tool for testing the proton structure and
perturbative Quantum Chromodynamics (pQCD). The dominant production
process is boson-gluon fusion between the incoming virtual photon and
a gluon in the proton. When the negative squared four momentum of
virtual photon, $Q^{2}$, is large compared to the proton mass, the
interaction is referred to as deep inelastic scattering (DIS).
Different kinematic variables which are used to describe $ep$ interactions
at HERA are: $Q^{2}$, the Bjorken scaling variable, $x$, and the
inelasticity, $y$.

In this analysis~\cite{abr11}, beauty production in DIS was studied using the
semileptonic decays to electrons. The measurements are compared to a
leading order plus parton shower Monte Carlo (\RAPGAP)~\cite{jung95} as well as QCD
predictions at next-to-leading order (NLO), calculated using HVQDIS
program~\cite{smith98}. This program is based on the fixed-flavour-number scheme
(FFNS), in which heavy flavours are generated dynamically in the hard
subprocess.

\section{Data Selection}
The analysis was performed with data taken by the ZEUS detector from
2004 to 2007, when HERA collided electrons or positrons with energy
$E_{e} = 27.5\xGeV{}$ with protons of energy $E_{p} = 920\xGeV{}$. The
corresponding integrated luminosity is 363\xipb{} at centre-of-mass
energy $\sqrt{s} = 318\xGeV{}$.  Standard cuts~\cite{thesis} were applied to select
DIS events in the range $Q^{2} > 10\xGeV^{2}$ and $0.05 < y <
0.7$. Electron candidates from semileptonic decays of $b$ quarks were
selected from Energy Flow Objects (EFOs) having a transverse momentum
$0.9 < p_{T}^{e} < 8\xGeV{}$ in the pseudorapidity range $|\eta^{e}| <
1.5$. Electrons from identified photon conversions were rejected. The
electron candidate was required to be associated to a jet with
$|\eta^{\text{jet}}| < 2.0$ and $p_{T}^{\text{jet}}$ > 2.5\xGeV{}.

\section{Signal Extraction}
For the identification of electrons from semileptonic $b$ decays,
variables sensitive to electron identification as well as to
semileptonic decays were used.  Electron identification uses the
measurement of the specific energy loss, $dE/dx$, in the central
tracking detector, the ratio of the energy deposited in the
calorimeter to the track momentum and the penetrating depth of the
energy deposited in the calorimeter.

\begin{wrapfloat}{ffig}[13]{r}{0.47\textwidth}
\vspace*{-20pt}
  \centering
  \hspace*{-1.5ex}\includegraphics[width=0.95\textwidth, clip=true]{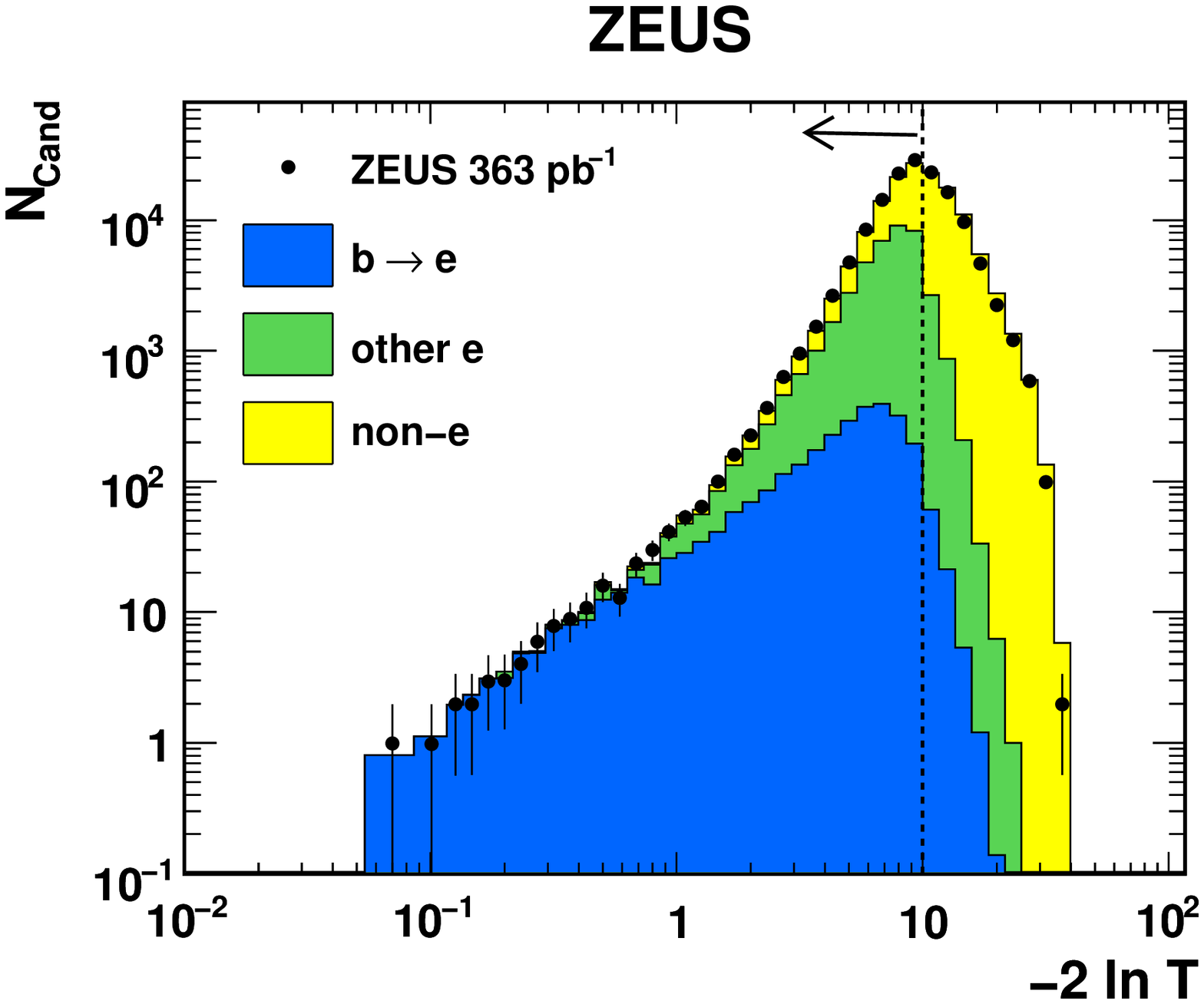}
  \vspace*{-10pt}\caption{The distribution of $-2 \ln T$, where T is the test
    function, using the beauty hypothesis for electron candidates.
  }
  \label{fig:likel_fit}    
\end{wrapfloat}
Semileptonic decays were separated from background using
$p_{T}^{\text{rel}}$, the relative transverse momentum component of
the electron candidate relative to the direction of jet axis; $\Delta
\phi$, the difference of the azimuthal angle between the electron
direction and the missing transverse momentum vector; and $d/\delta
d$, the significance of the reconstructed decay length, where $d$ is
defined as the distance in $X$-$Y$ between the secondary vertex and
the interaction point, projected onto the jet axis.  The six variables
were combined into one discriminating test-function variable using a
likelihood hypothesis.  For a given hypothesis of particle, $i$, and
source $j$, the likelihood, $\mathcal{L}_{ij}$, is given by
\begin{equation*}
\mathcal{L}_{ij} = \prod\limits_{l}\,\mathcal{P}_{ij}(d_{l})\,,
\end{equation*}
where $\mathcal{P}_{ij}(d_{l})$ is the probability to observe particle
$i$ from source $j$ with value $d_l$ of a discriminating variable. The
particle hypotheses $i \in \{e, \pi, K, p \}$ and the sources, $j$,
for electrons from semileptonic $b$ decays, electrons from other
sources including semileptonic $c$ decays and fake electrons
were considered. 
The test function $T_{ij}$ was defined as
 \begin{equation*}
T_{ij} = \frac{\alpha_{i} \alpha'_{j}\mathcal{L}_{ij}} {\sum
  \limits_{k, l} \alpha_{k} \alpha'_{l}\mathcal{L}_{kl}}
\end{equation*}
The $\alpha_{i}, \alpha'_{j}$ denote the prior probabilities taken
from Monte Carlo. 
The distribution of likelihood test function is shown in Figure~\ref{fig:likel_fit}. The
distribution is fit using the expected distributions for beauty, other
electrons and fake electrons to determine the fractions of events from
each source. The fit range of the test function was restricted to
$-2\ln T < 10$. The fit provides a very good description of data.
\begin{ffig}
  \hspace*{-0.4cm}\includegraphics[width=.31\textwidth]{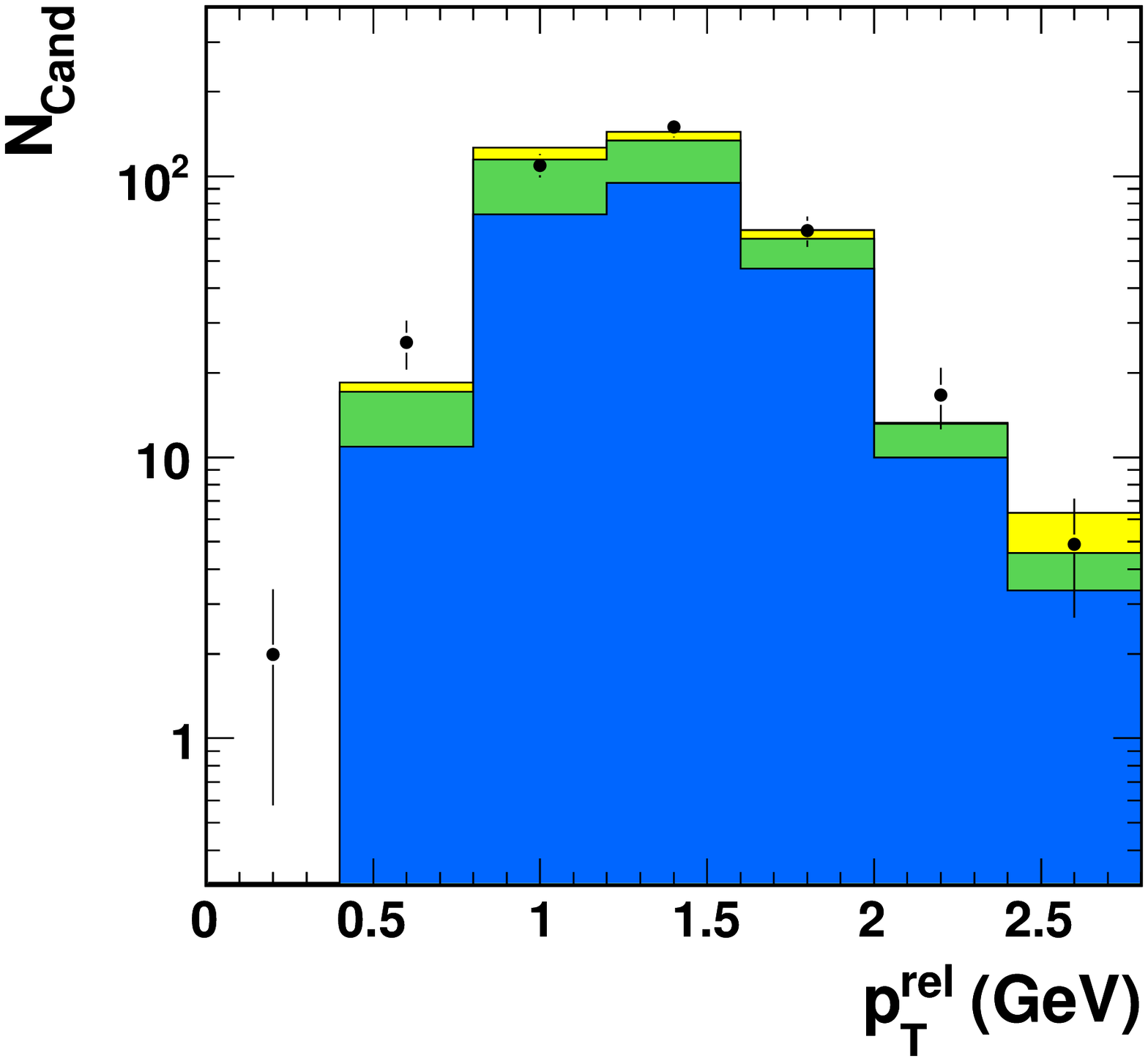}
  \hspace*{0.4cm}\includegraphics[width=.31\textwidth]{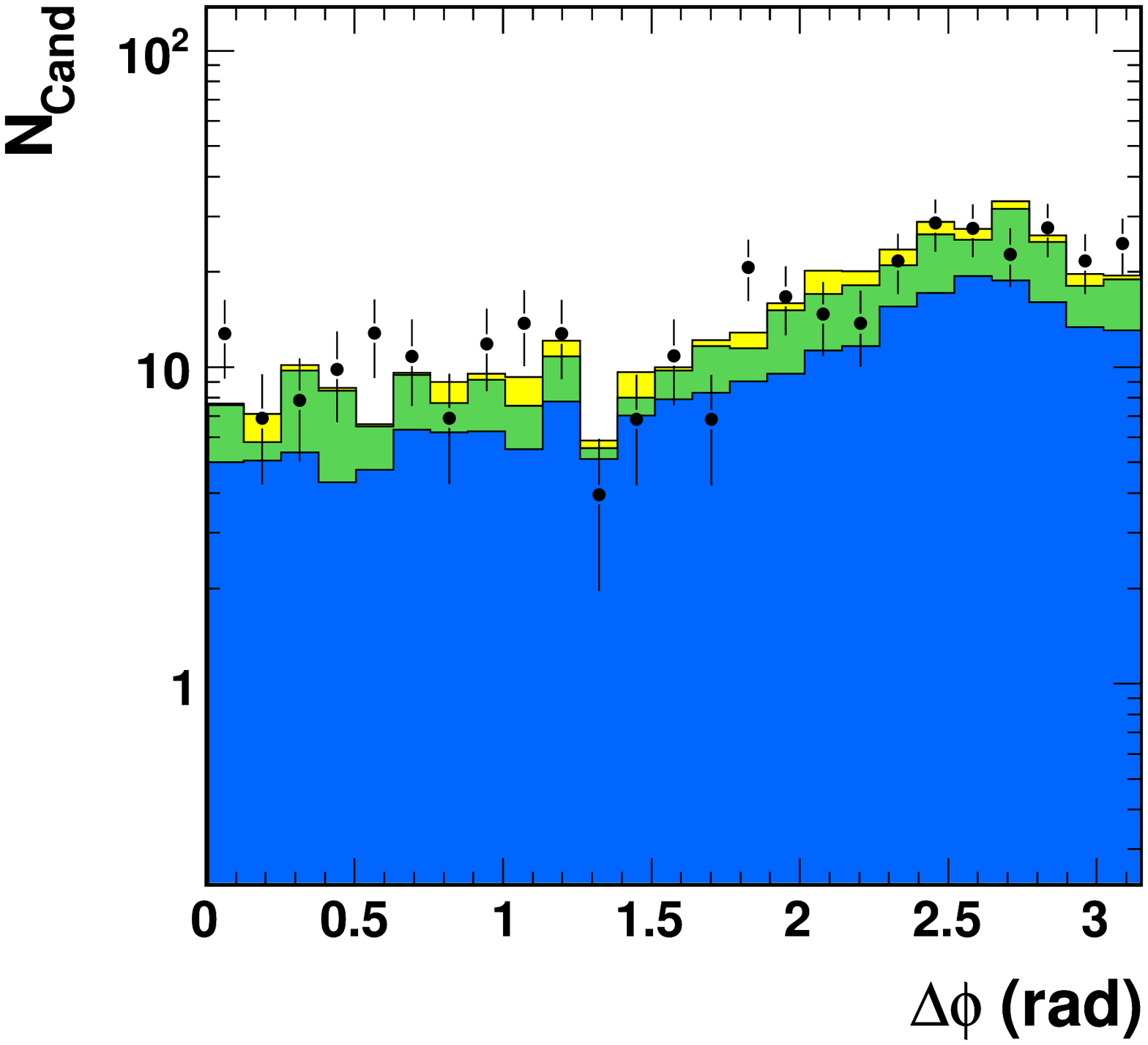}
  \hspace*{0.4cm}\includegraphics[width=.31\textwidth]{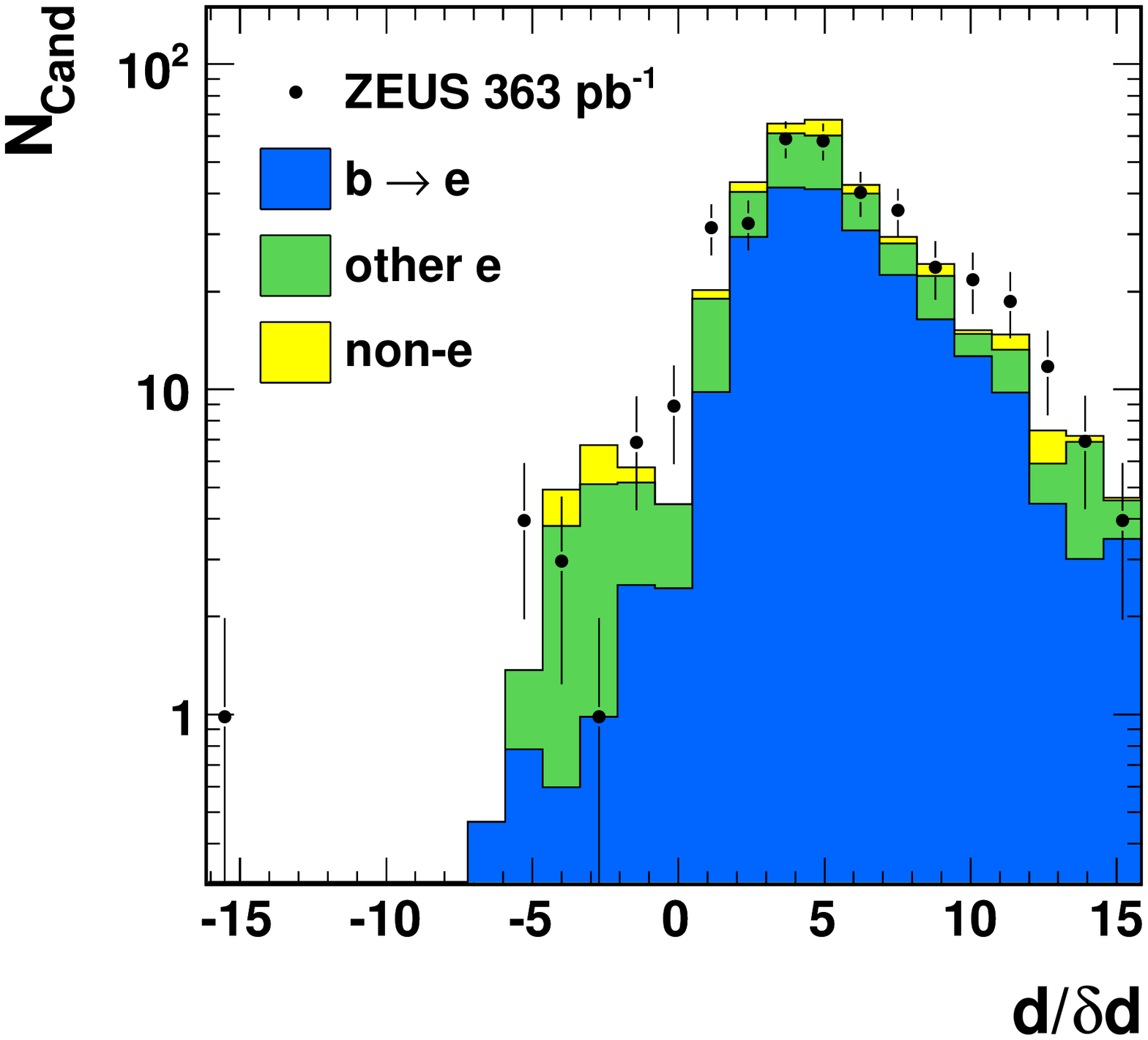}
  \vspace*{-2pt}
  {\caption{Distributions of $p_{T}^{\text{rel}}$, $\Delta
\phi$ and $d/\delta d$ for the signal-enriched region ($-2 \ln T < 1.5$).}
 \label{fig:ctrlplots}}
\end{ffig}
Figure~\ref{fig:ctrlplots} shows signal-enriched distributions ($-2
\ln T < 1.5$) for the variables in the likelihood-ratio test function,
which are sensitive to the different origins of the electron
candidates. All distributions are reasonably well described.

\section{Systematic Uncertainties}
The systematic uncertainties were calculated by varying the analysis
procedure and then repeating the fit to the likelihood
distributions~\cite{thesis}. Different sources of systematic uncertainty include
variation of the DIS selection, likelihood variables, different
background sources, jet energy scale, energy scales in the calorimeters and trigger
correction. No single dominant contribution was observed and the quadratic
sum of the systematic uncertainties was found to be of the same order
as the statistical uncertainty.

\section{Results}
The total visible cross section and differential cross sections for
$b$-quark production and the subsequent semileptonic decay to an
electron with $p_{T}^{e} > 0.9\xGeV$ in the range $|\eta^{e}| <$

\begin{wrapfloat}{ffig}[12]{r}{0.65\textwidth}
\vspace*{-5pt}
  \centering
  \includegraphics[width=0.97\textwidth, clip=true]{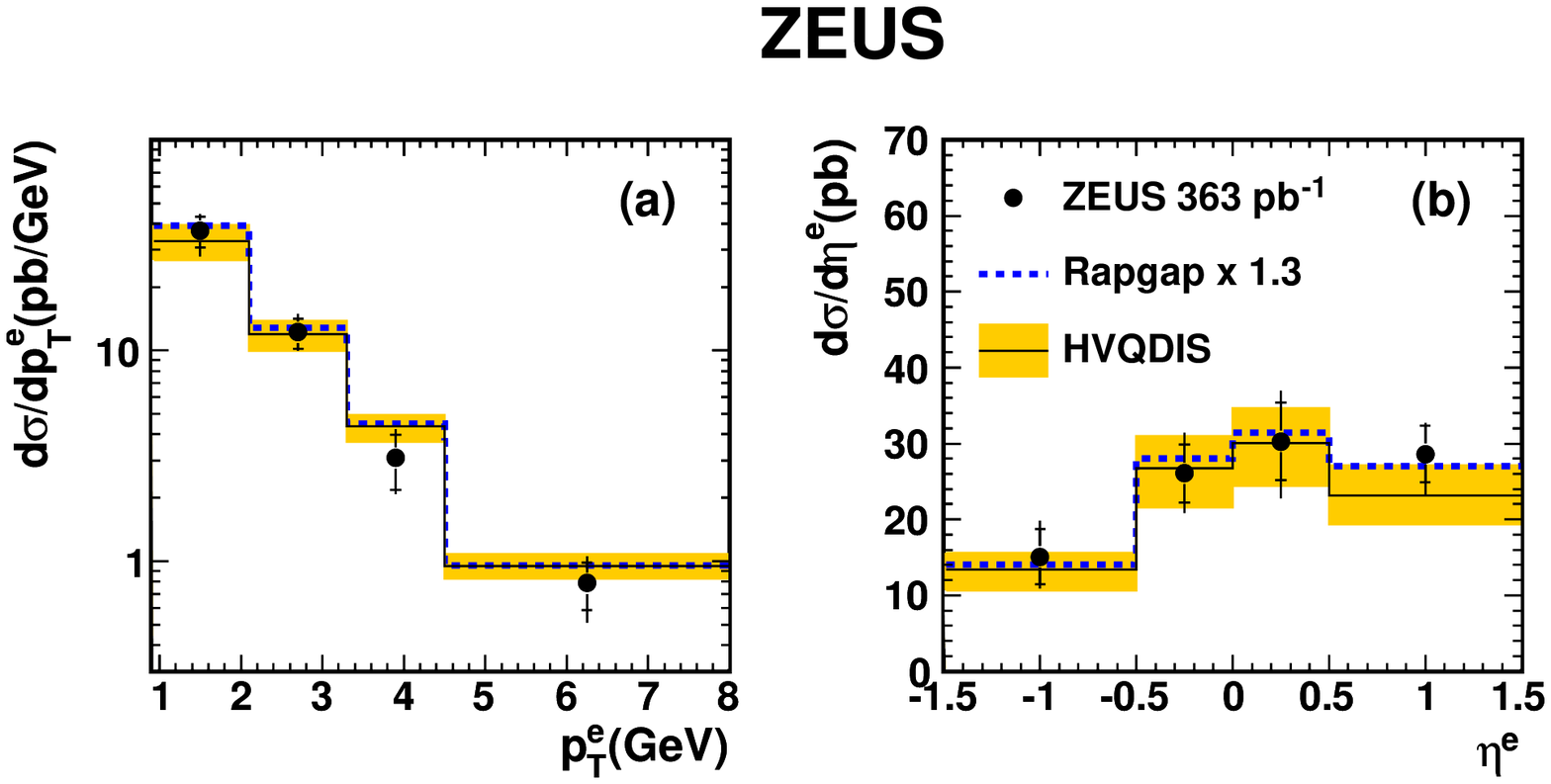}
  \vspace*{-2pt}\caption{Differential cross sections for electrons from $b$-quark
    decays as a function of (a) $p_{T}^{e}$ and (b)
    $\eta^{e}$.}
  \label{fig:diffxsec}
\end{wrapfloat}
\hspace*{-2.2ex}$1.5$ in DIS events with $Q^{2} > 10\xGeV^{2}$ and $0.05
< y < 0.7$ were measured. Figure~\ref{fig:diffxsec} shows differential
cross sections as a function of $p_{T}^{e}$ and $\eta^{e}$ compared to
the NLO QCD prediction and the \RAPGAP{} MC scaled to the data. Both
the descriptions from the NLO QCD calculation as well as the scaled
\RAPGAP{} cross sections describe the data well.

\section{Extraction of $F_{2}^{b\bar{b}}$}
The beauty contribution to the proton structure function $F_{2}$ can
be defined in terms of the inclusive double differential cross section
as a function of $x$ and $Q^{2}$,
\begin{equation*}
\frac{d^{2}\sigma^{b\bar{b}}}{dxdQ^{2}}=
\frac{2\pi\alpha^{2}}{xQ^{4}}\left(\left[1+(1-y)^{2}\right]F_{2}^{b\bar{b}}(x,Q^{2})-y^{2}F_{L}^{b\bar{b}}(x,Q^{2})\right) ,
\end{equation*}
where $F_{L}^{b\bar{b}}$ is the beauty contribution to the structure
function $F_{L}$.

\begin{wrapfloat}{ffig}[19]{r}{0.55\textwidth}
\vspace*{-25pt}
  \centering
  \includegraphics[width=0.92\textwidth, clip= true]{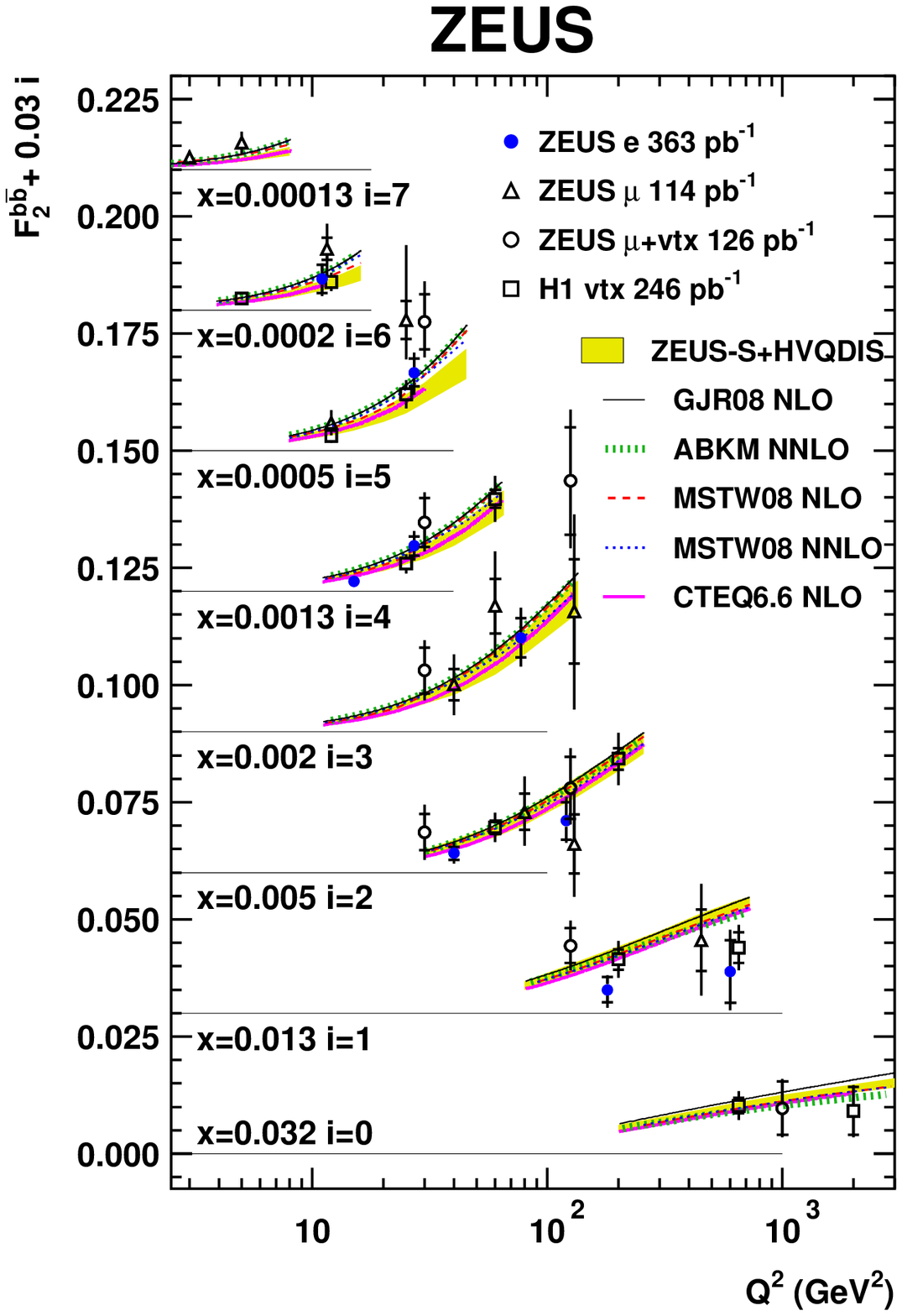}
  \vspace*{-10pt}\caption{$F_{2}^{b\bar{b}}$
    as a function of $Q^{2}$ for fixed $x$ values.}
  \label{fig:f2b}
\end{wrapfloat}
\hspace*{-2.2ex}The electron cross section, $\sigma_{b \shortrightarrow e}$, measured
in bins of $x$ and $Q^{2}$, was used to extract $F_{2}^{b\bar{b}}$ at
a reference point in the $x$-$Q^{2}$ plane using
\begin{equation*}
  F_{2}^{b\bar{b}}(x_{i},Q^{2}_{i}) = \frac{d^{2}\sigma_{b\shortrightarrow e}}{dxdQ^{2}} \cdot \frac{F_{2}^{b\bar{b},\text{NLO}}(x_{i},Q^{2}_{i})}{d^{2}\sigma_{b \shortrightarrow e}^{\text{NLO}}/dxdQ^{2}}
\end{equation*}
where $F_{2}^{b\bar{b},\text{NLO}}$ and $d^{2}\sigma_{b
  \shortrightarrow e}^{\text{NLO}}/dxdQ^{2}$ were calculated 
using the HVQDIS program.
Figure~\ref{fig:f2b} shows
$F_{2}^{b\bar{b}}$ as a function of $Q^{2}$ for fixed values of
$x$. The results from this measurement have been compared
with the previous measurements from the H1 and ZEUS collaborations. The different
measurements are consistent with each other. Also the results are
compared to several NLO and NNLO QCD predictions~\cite{theory}. The data are
reasonably well described by the different theory predictions.

\section{Summary}
A recent measurement of beauty production in DIS at HERA using decays
into electrons was presented. A likelihood-ratio test function was
used to identify the signal. The measured visible and differential
cross sections are in agreement with the NLO QCD
calculations. $F_{2}^{b\bar{b}}$ was extracted from the double
differential cross sections as a function of $x$ and $Q^{2}$, and is
in agreement with previous H1 and ZEUS measurements. For $Q^{2} >
10\xGeV^{2}$, this measurement represents the most precise
determination of $F_{2}^{b\bar{b}}$ by the ZEUS collaboration. The
results were also compared to several NLO and NNLO QCD calculations,
which provide a good description of the data.


\end{document}